\documentclass[prd,twocolumn,showpacs,preprintnumbers,superscriptaddress,amsmath]{revtex4}
\begin{document}
\title{Relationship between (2+1) and
(3+1)--Friedmann--Robertson--Walker cosmologies}
\author{Alberto Garc\'\i a}
\altaffiliation{aagarcia@fis.cinvestav.mx}
\affiliation{Departamento~de~F\'{\i}sica.
\\~Centro~de~Investigaci\'on~y~de~Estudios~Avanzados~del~IPN.\\
Apdo. Postal 14-740, 07000 M\'exico DF, MEXICO.\\}
\author{Mauricio Cataldo}
\altaffiliation{mcataldo@ubiobio.cl}
\affiliation{Departamento de F\'\i sica, Facultad de Ciencias,
Universidad del B\'\i o--B\'\i o, Avenida Collao 1202, Casilla
5-C, Concepci\'on, Chile.\\}
\author{Sergio del Campo}
\altaffiliation{sdelcamp@ucv.cl} \affiliation{Instituto de F\'\i
sica, Facultad de Ciencias B\'asicas y Matem\'aticas, Pontificia
Universidad Cat\'olica de Valpara\'\i so, Avenida Brasil 2950,
Valpara\'\i so, Chile.}
\date{\today}
\begin{abstract}
In this work we establish the correspondence between solutions to
the Friedmann--Robertson--Walker cosmologies for perfect fluid and
scalar field sources, where both ones fulfill state equations of
the form $p+\rho=\gamma f(\rho)$, not necessarily linear ones.
Such state equations are of common use in the case of
matter--fluids, nevertheless, for a scalar field, they introduce
relationships on the potential and kinetic scalar field energies
which restrict the set of solutions. A theorem on this respect is
demonstrated: From any given (3+1)--cosmological solution, obeying
the quoted state equations, one can derive its (2+1)--cosmological
counterpart or vice-versa. Some applications are given.
\end{abstract}
\maketitle
\section{Introduction}
Scalar fields play a crucial role in describing cosmological
models. In the standard big-bang theory such fields are included
for solving most of the problems found at very early times in the
evolution of the universe, and are called ``inflaton" scalar
field~\cite{Guth,Linde,Albrecht}. This scalar field is
characterized by its scalar potential. In fact, different
inflationary universe models differ from each other by its
potential form. For instance, the scalar potential related to new
inflation is very different from the potential of chaotic
inflation. These scalar fields not only are appropriated for
describing the evolution of the universe, but also they give the
necessary initial condition for the formation of the large scale
structure observed in the universe.

At the same time, the measurements of the luminosity--redshift
relations observed for the fifty newly discovered type Ia
supernovae with redshift $z > 0.35$~\cite{Perlmutter,Garnavich},
indicate that at present the universe is expanding with an
accelerated fashion suggesting a net negative pressure for the
universe. One of the plausible explanation of this astronomical
observation is based on the introduction of a scalar field, which
is called ``quintessence" or ``dark energy" scalar field. This
field has an associated effective scalar potential, which on its
turn plays a crucial role in describing tracker solutions.

Although these scalar fields are quite different in nature, there
are authors who think that the ``inflaton" and the ``quintessence"
fields might be of the same nature, in which a very specific
scalar potential form is used~\cite{Peebles}. In these cases, the
scalar field emerges in a kinetic--dominated regime at energy
densities above the tracker solution.

On the other hand, during the past decade the three--dimensional
gravity has been received much
attention~\cite{Teitelboim1,Carlip}. The reasons for this are many
and varied; however the principal one is the existence of black
holes solutions in (2+1)--anti de Sitter spacetimes, which possess
certain features inherent to the (3+1)--black
holes~\cite{Cataldo}. Often it is useful to consider a physical
system in lower dimensions, as for instance is done in quantum
field theory. Thus it is reasonable to extend this procedure to
gravity. It is believed that (2+1)--gravity will provide new
insights towards a better understanding of the physically relevant
(3+1)--gravity.

Most of the studies on this respect are related to the black hole
spacetimes. The literature on three--dimensional cosmological
models is rather scarce. Some Friedmann--Robertson--Walker
cosmologies (FRW) models had been analyzed in
Ref.~\cite{Giddings,Barrow,Cornish}. It is noteworthy to point out
that some of the non--trivial features of the (2+1)--gravity are
apparent in the behavior of cosmic strings and domain walls in
(3+1)--dimensions (see~\cite{Barrow} and the references therein).
Among the works on cosmology, one can cite~\cite{Cornish}, in
which Cornish and Frankel consider the three--dimensional Einstein
gravity and give the solutions for the isotropic dust--filled and
radiation--dominated universes for $k=-1,0,1$. They also study FRW
models in alternative relativistic theories of gravity. Cruz and
Mart\'\i nez~\cite{Cruz} have examined (2+1)--FRW models with a
perfect fluid and a homogeneous scalar field minimally coupled to
gravity.

The purpose of this contribution is to provide a new insight
towards the relation between (2+1) and (3+1)--standard cosmologies
coupled to a perfect fluid and a scalar field.

The outline of the present paper is as follows: In Sec. II we
briefly review the well known Einstein equations for the FRW
metrics in four and three--dimensional gravities coupled to a
perfect fluid and a scalar field. A theorem concerning the
correspondence of cosmological solutions of a certain kind is
formulated and demonstrated. In Sec III some applications are
given.

\section{FRW cosmology with a perfect fluid and a
scalar field}
In this section we shall consider homogeneous FRW
models filled with a perfect fluid and homogeneous scalar field
$\phi$ minimally coupled to gravity with a self--interacting
potential $V(\phi)$.

In (3+1)--dimensions the FRW model is given by the metric
\begin{equation}\label{fr42}
ds^2=dt^2-a(t)^2\left(\displaystyle\frac{dr^2}{1-kr^2}
+r^2d\Omega^2\right),
\end{equation}
where as usual $d\Omega^2:=d\theta^2+{\rm sin}^2\theta d\phi^2$,
$a(t)$ is the scale factor, and $k=-1,0,1$. The scale factor
$a(t)$ of the metric~(\ref{fr42}) is governed by equations,
commonly modelled in terms of perfect fluids and a cosmological
constant, if present, in which participate the matter energy
density $\rho_{_{4}}$, the matter isotropic pressure $p_{_{4}}$,
the conventional scalar field pressure $p_{_{\phi_{_{4}}}}$, the
scalar field energy density $\rho_{_{\phi_{_{4}}}}$, defined
through the scalar field $\phi_{_{4}}$, and self--interacting
potential $V(\phi_{_{4}})$ by
\begin{eqnarray}\label{eqsc47d}
\rho_{_{\phi_{_{4}}}}=\frac{1}{2} \dot{\phi}_{_{4}}^2+ V_{_{4}},
\,\,\,\,\, p_{_{\phi_{_{4}}}}=\frac{1}{2} \dot{\phi}_{_{4}}^2 -
V_{_{4}}.
\end{eqnarray}
The four--dimensional field equations are:
\begin{subequations}
\begin{eqnarray}\label{eqsc47a}
\displaystyle 3\frac{k+{\dot{a}}^2}{a^2}=\kappa_{_{4}} (
\rho_{_{4}}+\rho_{_{\phi_{_{4}}}}),
\end{eqnarray}
\begin{eqnarray}\label{eqsc47b}
\dot{\rho_{_{4}}}+3 \frac{\dot{a}}{a} (\rho_{_{4}}+p_{_4})=0,
\end{eqnarray}
\begin{eqnarray}\label{eqsc47c}
\dot\rho_{_{\phi_{_{4}}}}+3 \frac{\dot{a}}{a}
(\rho_{_{\phi_{_{4}}}}+p_{_{\phi_{_{4}}}})=0.
\end{eqnarray}
\end{subequations} The equation~(\ref{eqsc47b})
represents the conservation of matter content, while
Eq.~(\ref{eqsc47c}) corresponds to the energy conservation
associated to the scalar field. In this case the scalar field
interacts only with the gravitational field and hence, each
energy--momentum tensor is conserved independently from one to
another.

Similarly, in the (2+1)--dimensional case, the metric has the
form~(\ref{fr42}), where now $d\Omega^2:=d\theta^2$, and the
conventional scalar field pressure $p_{_{\phi_{_{3}}}}$ and the
scalar field energy density $\rho_{_{\phi_{_{3}}}}$ are defined
similarly to the Eq.~(\ref{eqsc47d})
\begin{eqnarray}\label{eqsc37d}
\rho_{_{\phi_{_{3}}}}=\frac{1}{2} \dot{\phi}_{_{3}}^2+ V_{_{3}},
\,\,\,\,\, p_{_{\phi_{_{3}}}}=\frac{1}{2} \dot{\phi}_{_{3}}^2 -
V_{_{3}}.
\end{eqnarray}

The following set of three--dimensional field equations has to be
fulfilled:
\begin{subequations}
\begin{eqnarray}\label{eqsc37a}
\displaystyle \frac{k+{\dot{a}}^2}{a^2}=\kappa_{_{3}} (
\rho_{_{3}}+\rho_{_{\phi_{_{3}}}}),
\end{eqnarray}
\begin{eqnarray}\label{eqsc37b}
\dot{\rho_{_{3}}}+2\frac{\dot{a}}{a} (\rho_{_{3}}+p_{_3})=0,
\end{eqnarray}
\begin{eqnarray}\label{eqsc37c}
\dot\rho_{_{\phi_{_{3}}}}+2 \frac{\dot{a}}{a}
(\rho_{_{\phi_{_{3}}}}+p_{_{\phi_{_{3}}}})=0.
\end{eqnarray}
\end{subequations} The main result of this
contribution can be stated in the form of a theorem
\newline {\bf Theorem}: Assuming invariance of the
time-coordinate, as well as the structural form invariance of the
scale factor $a(t)$ in both (2+1) and (3+1)--dimensional
cosmologies, coupled to a single scalar field and perfect fluid
subjected to state equations $p_{_{\phi}}+\rho_{_{\phi}}=\gamma
F(\rho_{_{\phi}})$ and $p+\rho=\gamma f(\rho)$ respectively, where
$F(\rho_{_{\phi}})$ and $f(\rho)$ are dimensional invariant
structural functions, then the gravitational constant $\kappa$ and
the state parameters $\gamma$ and $\Gamma$ obey the following
rules:
\begin{eqnarray}\label{eqsct1}
\frac{\kappa_{_{4}}}{3}\rightleftharpoons\kappa_{_{3}},\,\,3
\gamma_{_{4}}\rightleftharpoons 2\gamma_{_{3}},\,\,\,\,
3\Gamma_{_{4}}\rightleftharpoons
2\Gamma_{_{3}},\,\,\,\, 
\end{eqnarray}
while the structural functions fulfill:
\begin{eqnarray}\label{eqsct155}
\rho_{_{4}}\rightleftharpoons\rho_{_{3}},\,\,
\rho_{_{\phi_{_{4}}}}\rightleftharpoons\rho_{_{\phi_{_{3}}}},
\sqrt{\frac{3}{2}}{\phi}_{_{4}}\rightleftharpoons{\phi}_{_{3}},\,\,
\nonumber\\
V_{_{4}}-\frac{1}{4}{{\dot\phi}_{_{4}}}^2\rightarrow
V_{_{3}},\,\,\,
V_{_{3}}+\frac{1}{6}{{\dot\phi}_{_{3}}}^2\rightarrow V_{_{4}}.
\end{eqnarray}
{\bf Proof}: Considering that the time coordinate $t$ as well the
scale factor $a(t)$ remain unchanged, then from
Eqs.~(\ref{eqsc47a}) and~(\ref{eqsc37a}), assuming the densities
$\rho$ and $\rho_{_{\phi}}$ independent, one has
\begin{eqnarray}\label{veqsct8}
&&\frac{{\dot{a}}^2+k}{a^2}=\frac{\kappa_{_{4}}}{3}(\rho_{_{4}}+\rho_{_{\phi_{_{4}}}})
=\kappa_{_{3}}(\rho_{_{4}}+\rho_{_{\phi_{_{3}}}})\Rightarrow
\rho_{_{4}}\rightleftharpoons\rho_{_{3}},\nonumber\\
&&\rho_{_{\phi_{_{4}}}}\rightleftharpoons\rho_{_{\phi_{_{3}}}},
\frac{\kappa_{_{4}}}{3}\rightleftharpoons\kappa_{_{3}}.
\end{eqnarray}
At this point, it is in order to explain what we mean by a
dimensional invariant structural function; as such we define a
function whose dependence on the $t$--coordinate or scalar
function $\phi$--variable, is the same in both dimensions and, if
gravitational and state parameters are present in it,
under~(\ref{eqsct1}) the form of the function remains unchanged.

Further, assuming that the state equation for matter is of the
form $p+\rho=\gamma f(\rho)$, where $f(\rho)$ is a structurally
invariant function, matter conservation equations yield
\begin{eqnarray}\label{veqsct7}
\frac{da}{a}=-\frac{1}{3\gamma_{_{4}}}\frac{d\rho_{_{4}}}{f(\rho_{_{4}})}=-\frac{1}{2
\gamma_{_{3}}}\frac{d\rho_{_{3}}}{f(\rho_{_{3}})},
\end{eqnarray}
hence, because of by assumption $f(\rho_{_{3}})\rightleftharpoons
f(\rho_{_{4}})$, one has
\begin{eqnarray}\label{veqsct78}
\ln{\frac{a}{a_0}}=-\frac{1}{3\gamma_{_{4}}}\int^{\rho_{_{4}}}{\frac{d\rho}{f(\rho)}}
=-\frac{1}{2\gamma_{_{3}}}\int^{\rho_{_{3}}}{\frac{d\rho}{f(\rho)}},
\end{eqnarray}
therefore
\begin{eqnarray}\label{veqsc641215}
3\gamma_{_{4}}\rightleftharpoons 2\gamma_{_{3}}.
\end{eqnarray}
Next, assuming that the state equation for the scalar field is of
the form $p_{_{\phi}}+\rho_{_{\phi}}=\Gamma F(\rho_{_{\phi}})$,
where $F(\rho_{_{\phi}})$ is a structurally invariant function,
the scalar field conservation equations yield
\begin{eqnarray}\label{veqsct71}
\frac{da}{a}=-\frac{1}{3\Gamma_{_{4}}}\frac{d\rho_{_{\phi_{_{4}}}}}{F(\rho_{_{\phi_{_{4}}}})}=
-\frac{1}{2\Gamma_{_{3}}}\frac{d\rho_{_{\phi_{_{3}}}}}{F(\rho_{_{\phi_{_{3}}}})},
\end{eqnarray}
hence,
\begin{eqnarray}\label{veqsct72}
\ln{\frac{a}{a_0}}=-\frac{1}{3\Gamma_{_{4}}}\int^{\rho_{_{\phi_{_{4}}}}}
{\frac{d\rho_{_{\phi}}}{F(\rho_{_{\phi}})}}
=-\frac{1}{2\Gamma_{_{3}}}\int^{\rho_{_{\phi_{_{3}}}}}
{\frac{d\rho_{_{\phi}}}{F(\rho_{_{\phi}})}},
\end{eqnarray}
therefore
\begin{eqnarray}\label{eqsc642}
3\Gamma_{_{4}}\rightleftharpoons 2\Gamma_{_{3}}.
\end{eqnarray}
From  the equations $ {\dot{\phi}}^2=\rho_{_{\phi}}+p_{_{\phi}}$,
one obtains
\begin{eqnarray}\label{eqsc643}
{\dot{\phi_{_{4}}}}^2=\Gamma_{_{4}}F(\rho_{_{\phi}}),\,\,{\dot{\phi_{_{3}}}}^2=\Gamma_{_{3}}F(\rho_{_{\phi}}),
\end{eqnarray}
thus
\begin{eqnarray}\label{eqsc644}
\frac{\dot{\phi_{_{4}}}}{\sqrt{\Gamma_{_{4}}}}=\frac{\dot{\phi_{_{3}}}}{\sqrt{\Gamma_{_{3}}}}=\sqrt{F(\rho_{_{\phi}})},
\end{eqnarray}
consequently, taking into account Eqs.~(\ref{eqsc642}), up to
additive constants one gets
\begin{eqnarray}\label{eqsc646}
\dot{\phi_{_{4}}}\rightleftharpoons\sqrt{\frac{2}{3}}\dot{\phi_{_{3}}}\rightleftharpoons
\phi_{_{4}}\rightleftharpoons\sqrt{\frac{2}{3}}\phi_{_{3}}.
\end{eqnarray}
Finally, from equations $2
V(\phi)=\rho_{_{\phi}}-p_{_{\phi}}=2\rho_{_{\phi}}-\Gamma
F(\rho_{_{\phi}})$, namely
\begin{eqnarray}\label{eqsc6469}
V(\phi_{_{4}})=\rho_{_{\phi_{_{4}}}}-\frac{\Gamma_{_{4}}}{2}
F(\rho_{_{\phi}}), \,\, V(\phi_{_{3}})=
\rho_{_{\phi_{_{3}}}}-\frac{\Gamma_{_{3}}}{2} F(\rho_{_{\phi}}),
\end{eqnarray}
one arrives at
\begin{eqnarray}\label{eqsc10}
V_{_{4}}-\frac{1}{4}{{\dot\phi}_{_{4}}}^2\rightarrow V_{_{3}},
V_{_{3}}+\frac{1}{6}{{\dot\phi}_{_{3}}}^2\rightarrow V_{_{4}}.
\end{eqnarray}
It is noteworthy to point out that this theorem is based mainly on
the equality of the dynamical structure of the field equations in
different dimensions. The physical content of solutions in the
presence of perfect fluids changes as viewed from different
dimensional spacetimes; for instance, starting in (3+1)--cosmology
with dust, the (2+1)--counterpart will be a fluid with
$\gamma_{_{3}}=1/2$, therefore is no way within this treatment to
relate dust with dust in the considered dimensions. On the other
hand, one has to recall the constraints that energy conditions
impose on the matter.

{\bf Corollary}: Cosmologies in (2+1) and (3+1)--dimensions
coupled to a single scalar field, in which is assumed the
time--coordinate as well as the scale factors to be the same, up
to the parametrization of the gravitational and state constants
given below, for both (3+1) and (2+1)--spaces, are related
according to the following rules:
\begin{eqnarray}\label{eqsct12}
&&\frac{\kappa_{_{4}}}{3}\rightleftharpoons\kappa_{_{3}},\,\,
3\Gamma_{_{4}}\rightleftharpoons 2\Gamma_{_{3}},\,\,\,\, \nonumber
\\ &&\rho_{_{\phi_{_{4}}}}\rightleftharpoons\rho_{_{\phi_{_{3}}}},
\sqrt{\frac{3}{2}}{\phi}_{_{4}}\rightleftharpoons{\phi}_{_{3}},
\nonumber \\ 
&&V_{_{4}}-\frac{1}{4}{{\dot\phi}_{_{4}}}^2\rightarrow
V_{_{3}},\,V_{_{3}}+\frac{1}{6}{{\dot\phi}_{_{3}}}^2\rightarrow
V_{_{4}}.
\end{eqnarray}
The proof follows immediately from the theorem above.

Remark: It is noteworthy to point out that state equations of the
form $p+\rho=\gamma (\rho)$ although they work well for
matter-perfect fluids, in the case of a scalar field they
introduce a relation between the scalar potential $V(\phi)$ and
the kinetic energy $\dot \phi^2/2$, restricting in this way the
set of scalar field solutions. It would be of interest to search
for wider classes of scalar field solutions and establish an
algorithm to relate FRW cosmological models in different
dimensions, this kind of research is in progress.

\section{Generating solutions via transformations}
From any given solution in (2+1)--cosmology with a single scalar
field, the above relations~(\ref{eqsct1}) or~(\ref{eqsct12}) allow
one to construct solutions of the similar kind in
(3+1)--spacetime, and conversely. In particular, in what follows
we shall restrict our study to flat FRW cosmologies.

\subsection{Barrow--Burd--Lancaster and Madsen
inflationary solutions}
Barrow, Burd and Lancaster~\cite{Barrow}
determined two exact solutions exhibiting the evolution of
cosmological models containing self--interacting scalar fields
with physically interesting potentials in the zero-curvature FRW
model. In this case the equation of state, a non--linear one, of
the scalar field is given by
$p_{_{\phi_{3}}}+\rho_{_{\phi_{3}}}=\alpha
\rho_{_{\phi_{3}}}^{1/2}$. One of the solutions is given by
\begin{subequations}
\label{eq:whole}
\begin{equation}
a(t)=a_{_{0}} \, e^{-\frac{1}{4}(\kappa_{_{3}} A^2 \, e^{\pm
\sqrt{8 \mu} \, t})}, \label{subeq:1}
\end{equation}
\begin{equation}
\Phi_{_{3}}(t)=A e^{\pm \sqrt{8 \mu} \, t/2},
\end{equation}
\begin{equation}
V(\phi_{_{3}})= \mu \left(\frac{1}{2} \kappa_{_{3}}
\phi_{_{3}}^4-\phi_{_{3}}^2 \right),
\end{equation}
\end{subequations}
where $a_{_{0}}$, $\mu$ and $A$ are constants.

Thus using the relations~(\ref{eqsct12}), we can obtain for the
(3+1)--counterpart the following zero--curvature scalar FRW
cosmology:
\begin{subequations}
\begin{eqnarray}
a(t)=a_{_{0}} \, e^{-\frac{1}{4}(\frac{\kappa_{_{4}}}{3} A^2 \,
e^{\pm \sqrt{8 \mu} \, t})},
\end{eqnarray}
\begin{eqnarray}
\phi_{_{4}}(t)=A \sqrt{\frac{2}{3}} \, e^{\pm \sqrt{8 \mu} \,
t/2}.
\end{eqnarray}
\begin{eqnarray} V(\phi_{_{4}})= \mu \left(\frac{3}{8}
\kappa_{_{4}} \phi_{_{4}}^4-\phi_{_{4}}^2 \right),
\end{eqnarray}
\end{subequations} This four--dimensional cosmological
model has been previously found by Madsen~\cite{Madsen}. In this
case the inflationary solution corresponds to the negative sign in
the exponents and admits symmetry breaking.

A second four--dimensional inflationary solution may be obtained
from the (2+1)--expanding universe given by
\begin{subequations}
\begin{eqnarray}\label{qw}
a(t)=t^2 \sqrt{1+\frac{A}{t^3}},
\end{eqnarray}
\begin{eqnarray}\label{qw1}
\phi_{_{3}}=\sqrt{2 \rho_{_{0}}} \, \ln \left[ C_{_{0}} t^2 \left
(1+\frac{A}{t^3} \right) \right],
\end{eqnarray}
\begin{eqnarray}\label{qw2}
V(\phi_{_{3}})= 12 \rho_{_{0}} \frac{t}{A+t^3}= 12 \rho_{_{0}}
C_{_{0}} \, e^{-\Phi_{_{3}}/\sqrt{2 \rho_{_{0}}}},
\end{eqnarray}
\end{subequations} where $A$, $\rho_{_{0}}$ and
$C_{_{0}}$ are constants. For writing the above solution in the
Barrow's form one has to choose $\rho_{_{0}}=\frac{1}{4
\kappa_{_{3}}}$ and $C_{_{0}}=\frac{\kappa_{_{3}} \Lambda}{3}$.

Thus, using the relations~(\ref{eqsct12}) we can obtain for the
(3+1)--counterpart of~(\ref{qw})--(\ref{qw2}) the following
zero--curvature scalar FRW cosmology:
\begin{subequations}
\begin{eqnarray}
a(t)=t^2 \sqrt{1+\frac{A}{t^3}},
\end{eqnarray}
\begin{eqnarray}
\phi_{_{4}}=\sqrt{\frac {4 \rho_{_{0}}}{3}} \, \ln \left[ C_{_{0}}
t^2 \left (1+\frac{A}{t^3} \right) \right],
\end{eqnarray}
\begin{eqnarray}
V(\phi_{_{4}})= \frac{\rho_{_{0}}}{3} \frac{40 t^6+32 A
t^3+A^2}{t^2(A+t^3)^2}.
\end{eqnarray}
\end{subequations}
This solution, as far as we know, has not been reported before in
the literature. Then it is of certain interest to study this
four--dimensional inflationary cosmological model.

\subsection{Four--dimensional version of the
Cruz--Mart\'\i nez solution} Cruz and Mart\'\i nez~\cite{Cruz}
have obtained a solution which describes a (2+1)--flat FRW
cosmology with a homogeneous scalar field with a self--interacting
potential whose energy density redshifts as $a^{-2 \gamma}$, where
$a(t)$ is the scale factor. The solution with state equation
$p_{_{\phi_{_{3}}}}+\rho_{_{\phi{_{3}}}} = \gamma_{_{3}}
\rho_{_{\phi{_{3}}}}$ may be written as:
\begin{subequations}
\begin{eqnarray}
a(t)=\left(t_{_{0}}+\epsilon_{_{a}}
\gamma_{_{3}}\sqrt{\kappa_{_{3}}}\,t \right )^{1/\gamma_{_{3}}},
\end{eqnarray}
\begin{eqnarray}
\phi_{_{3}}(t)=\frac{1}{\sqrt{\kappa_{_{3}}\gamma_{_{3}}}} \, \ln
\left(t_{_{0}}+\epsilon_{_{a}}
\gamma_{_{3}}\sqrt{\kappa_{_{3}}}\,t \right )+\phi_{_{0}},
\end{eqnarray}
\begin{eqnarray}
V(\phi_{_{3}})=\frac{2-\gamma_{_{3}}}{2} \,
e^{-2\sqrt{\kappa_{_{3}} \gamma{_{3}}}(\phi_{_{3}} -
\phi_{_{0}})},
\end{eqnarray}
\end{subequations} where $\epsilon_{_{a}}=\pm 1$,
$\rho_{_{\phi_{_{0}}}}$, $\phi_{_{0}}$ and $a_0$ are constants of
integration.

Using the relations~(\ref{eqsct12}) we derive the (3+1)--scalar
FRW cosmology counterpart:
\begin{subequations}
\begin{eqnarray}\label{28}
a(t)=\left(t_{_{0}}+\frac{3\gamma_{_{4}}\epsilon_{_{a}}}{2}\sqrt{\frac{\kappa_{_{4}}}{3}}\,t
\right)^{\frac{2}{3 \gamma_{_{4}}}},
\end{eqnarray}
\begin{eqnarray}\label{281}
\phi_{_{4}}(t)= \frac{2}{\sqrt{3 \kappa_{_{4}} \gamma_{_{4}}}} \,
\ln
\left(t_{_{0}}+\frac{3\gamma_{_{4}}\epsilon_{_{a}}}{2}\sqrt{\frac{\kappa_{_{4}}}{3}}\,t
\right )+\phi_{_{0}},
\end{eqnarray}
\begin{eqnarray}\label{282}
V(\phi_{_{4}})=\frac{(2-\gamma_{_{4}})}{2}\,
e^{-\sqrt{3\kappa_{_{4}}\gamma_{_{4}}}(\phi_{_{4}}-\phi_{_{0}})},
\end{eqnarray}
\end{subequations}
where $\epsilon_{_{a}}=\pm 1$, $\rho_{_{\phi_{_{0}}}}$,
$\phi_{_{0}}$ and $a_0$ are the same constants of integration as
for~(\ref{28})--(\ref{282}). Notice that for $3 \gamma_{_{4}} < 2$
this solution describes an accelerating universe.

\subsection{Barrow-Saich solution}
The previously treated  solutions contain only a single scalar
field as a source. We shall now assume the source of a
(3+1)--universe to be a perfect fluid, with a equation of state
$p_{_{4}}+\rho_{_{4}}=\gamma_{_{4}} \, \rho_{_{4}}$ and a scalar
field with equation of state
$p_{_{\phi_{4}}}+\rho_{_{\phi_{4}}}=\frac{\gamma_{_{4}}}{2} \,
\rho_{_{\phi_{4}}}$. The conservation equations of the perfect
fluid and scalar field give $\rho_{_{4}}=A_{_{4}} a^{-2 /(3
\gamma_{_{4}})}$ and $\rho_{_{\phi_{4}}}=A_{_{\phi_{4}}} a^{- /(3
\gamma_{_{4}})}$ respectively. In this case the obtained solution
is defined through~\cite{Saich}:
\begin{subequations}
\begin{eqnarray}
a(t)=\left(\frac{3 \kappa_{_{4}}}{16} \gamma^2_{_{4}} A_{_{\phi}}
(t-t_{0})^2-A/A_{_{\phi}} \right)^{2/(3 \gamma_{_{4}})},
\end{eqnarray}
\begin{eqnarray}
\phi_{_{4}}(t)=\phi_{0}+  \sqrt{\frac{2}{3}}
\sqrt{\frac{1}{\gamma_{_{4}} \kappa_{_{4}}}} \, \ln \left[\frac{3
\gamma_{_{4}}}{4} A_{_{\phi}} \sqrt{\frac{\kappa_{4}}{3}}
(t-t_{0}) \right.
\nonumber \\
\left.   + \sqrt{\frac{ 3 \gamma^2_{_{4}}}{16} \kappa_{_{4}}
A^2_{_{\phi}} (t-t_{0})^2-A} \right], \hspace{2cm}
\end{eqnarray}
\begin{eqnarray}
V(\phi_{_{4}}) = \frac{1}{2}(4-\gamma_{_{4}}) A^2_{_{\phi}}
\frac{{e^{\sqrt{3 \kappa_{4} \gamma_{_{4}}/2} \,
(\phi_{_{4}}-\phi_{0})}}}{[e^{\sqrt{3 \kappa_{4} \gamma_{_{4}}/2}
\, (\phi_{_{4}}-\phi_{0})}-A]^2},
\end{eqnarray}
\end{subequations}
where $t_0$, $A_{_{\phi}}$, $\phi_{_{0}}$ and $A$ are arbitrary
constants. Again, in order to have an accelerating universe we
should restrict $\gamma_{_{4}}< 4/3$.

Now using the relations~(\ref{eqsct1}) and~(\ref{eqsct155}) we can
obtain the following (2+1)--flat FRW cosmology:
\begin{subequations}
\begin{eqnarray}
a(t)=\left(  \frac{\kappa_{_{3}}}{4} \gamma^2_{_{3}} A_{_{\phi}}
(t-t_{0})^2-A/A_{_{\phi}} \right)^{1/\gamma_{_{3}}},
\end{eqnarray}
\begin{eqnarray}
\phi_{_{3}}(t)= \phi_{0} + \sqrt{\frac{2}{\gamma_{_{3}}
\kappa_{_{3}}}} \, \ln \left[ \frac{\gamma_{_{3}}}{2} A_{_{\phi}}
\sqrt{\kappa_{3}} (t-t_{0})  \right.
\nonumber \\
\left. + \sqrt{\frac{\gamma^2_{_{3}}}{4} \kappa_{_{3}}
A^2_{_{\phi}} (t-t_{0})^2-A} \right]
\end{eqnarray}
\begin{eqnarray}
V(\phi_{_{3}})= (4-\gamma_{_{3}}) A^2_{_{\phi}} \frac{{e^{\sqrt{2
\kappa_{3} \gamma_{_{3}}} (\phi_{_{3}}-\phi_{0})}}}{[e^{\sqrt{2
\kappa_{3} \gamma_{_{3}}} (\phi_{_{3}}-\phi_{0})}-A]^2},
\end{eqnarray}
\end{subequations} where $t_0$, $A_{_{\phi}}$,
$\phi_{_{0}}$ and $A$ are constants.

\section{Concluding Remarks}
In this report it is established a theorem which allows one to put
in correspondence (2+1) and (3+1)--FRW cosmologies. The
established relationship holds for solutions modelled through
conventional perfect fluids--matter and scalar potential
fluids--obeying state equations of the form $p+\rho=\gamma
f(\rho)$, notice that they are not necessarily linear state
equations, which in the case of the scalar field impose conditions
on the energies and consequently restrict the set of scalar field
solutions. A procedure to derive from a given (2+1)--FRW solution
a (3+1)--FRW solution and vice--versa was exhibited. For instance,
we have shown that the Barrow et al (2+1)--metric becomes the
(3+1)--Madsen cosmological solution. Moreover, from the
(3+1)--Barrow--Saich metric structure one derive its
(2+1)--counterpart. Information about the physical interpretation
of the above considered solutions can be found in the quoted
references. One of the important features of this approach resides
on the possibility of interpreting the related cosmologies from
dimensionally different points of view; state equations in
different dimensions reveal dissimilar physical content. For
instance, (3+1)--radiation possesses as counterpart the
(2+1)--stiff matter. In spite of the generality of this theorem,
based mainly on the equality of the dynamical structure in
different dimensions, there are classes of state equations which
remain uncovered by this theorem, for example the polytropic law;
a work on this line is in progress and will be published
elsewhere.

\section{acknowledgements}
AG acknowledges the hospitality of the Physics Department of
Universidad del B\'\i o--B\'\i o where a considerable part of this
work was done. This work was partially supported by CONICYT
through grants N$^0$ 7010485 (MC and AG), FONDECYT N$^0$ 1010485
(MC and SdC), N$^0$ 1030469 (SdC and MC); CONACYT N$^0$ 38495--E
(AG) and CONICYT/CONACYT N$^0$ 2001-5-02-159. It also was
supported by the Direcci\'on de Investigaci\'on de la Universidad
del B\'\i o--B\'\i o (MC) and by grant 123.764/2003 of
Vicerrectoria de Investigaci\'on y Estudios Avanzados of
Pontificia Universidad Cat\'olica de Valpara\'{\i}so (SdC).

\end{document}